\documentclass[preprint]{aastex62}

\shorttitle{VY CMa Emission Spectrum}
\shortauthors{Humphreys et al.}
\begin{document}

\title{The Unexpected  Spectrum of the Innermost Ejecta of the Red Hypergiant
VY CMa\footnote{Based on observations made with the NASA/ESA Hubble Space Telescope, which is operated by the Association of Universities for Research in
Astronomy, Inc., under NASA contract NAS 5-26555.}}

\correspondingauthor{Roberta Humphreys}
\email{roberta@umn.edu, lziurys@email.arizona.edu}

\author{Roberta M. Humphreys}
\affiliation{Minnesota Institute for Astrophysics,  
University of Minnesota,
Minneapolis, MN 55455, USA}
\nocollaboration

\author{L. M. Ziurys}
\affiliation{Departments of Astronomy and Chemistry,
University of Arizona}
\nocollaboration

\author{J. J. Bernal}
\affiliation{Department of Chemistry, University of Arizona}
\nocollaboration

\author{Michael S. Gordon}
\affiliation{USRA-SOFIA Science Center, NASA Ames Research Center,
Moffett Field, CA 94035, USA }
\nocollaboration

\author{L. Andrew Helton}
\affiliation{USRA-SOFIA Science Center, NASA Ames Research Center,
Moffett Field, CA 94035, USA }
\nocollaboration

\author{Kazunori Ishibashi}
\affiliation{Graduate School of Science, Nagoya University, Nagoya, 464-8602, Japan}
\nocollaboration

\author{Terry J. Jones}
\affiliation{Minnesota Institute for Astrophysics,
University of Minnesota,
Minneapolis, MN 55455, USA}
\nocollaboration

\author{A. M. S. Richards}
\affiliation{Jodrell Bank,
Department of Physics and Astronomy, University of Manchester, UK}

\author{Wouter Vlemmings}
\affiliation{Department of Space, Earth and Environment, Chalmers University of Technology, Onsala Space Observatory, 439 92, Onsala, Sweden}
\nocollaboration

%% Mark off the abstract in the ``abstract'' environment. 
\begin{abstract}
{\it HST/STIS} spectra of the small clumps and filaments closest to the 
central star in VY CMa reveal that the very strong K I emission and TiO and VO molecular emission,  long thought to form in a dusty circumstellar shell, actually originate in  a few small clumps  100's of AU from the star. The K I lines are 10 to 20 times stronger in these nearest ejecta than on the star.  The observations also confirm VO as a  circumstellar molecule.  In this letter we discuss the  spectra of the features, their motions and ages, and the identification of the molecular emission. The strength of the atomic and molecular features in the small clumps 
present an astrophysical problem for the excitation process. We show that the clumps must have a nearly clear line of sight to the star's radiation.  
\end{abstract}


\section{Introduction} \label{sec:intro}

The red hypergiant VY CMa is the most extreme known case of stellar 
surface activity in terms of energy, mass outflow, and size scale. 
It is one of the strongest 10--20 $\mu$m sources in the sky, with  
powerful maser activity, an unusual set of molecular features,  
the strongest known \ion{K}{1} emission, and other rare characteristics. 
This unique combination of traits makes it one of the most important 
objects for understanding episodic mass loss and stellar activity.  Unfortunately,    
ground-based spectra do not resolve the dust-formation zone and 
the geometry of the inner ejecta.  In this paper we report    
{\it HST/STIS} spectroscopy  with spatial resolution 0\farcs1.  
Unexpectedly, these data reveal that molecular and \ion{K}{1} 
emission are concentrated in a few  bright 
knots, $\approx$  0.3\arcsec from the star, rather than being distributed in a dusty circumstellar envelope.

Earlier multi-wavelength {\it HST/WFPC2} images revealed a circumstellar environment
with numerous knots, arcs and prominence-like loops visible by scattered light
\citep{Smith}. Using transverse motions measured from second epoch images plus
ground-based Doppler velocities (Humphreys et al. 2005, Paper I), we mapped the 3D morphology of the ejecta and showed that these features were both spatially and 
kinematically distinct from the surrounding nebulosity (Humphreys et al. 2007 (PaperII), Jones et al. 2007). 
They were ejected at different times over several hundred years in different directions, apparently by localized processes from different regions on the 
star.

Mass estimates from surface photometry in the HST images and from near and mid-IR imaging and polarimetry of the SW Clump, yield minimum masses (dust + gas) of $>$ 5 $\times$ 10$^{-3}$ M$_{\odot}$ \citep{Shenoy,Gordon}. ALMA sub-millimeter 
observations 
 reveal prominent dust components $\approx$ 0\farcs5 from the star with the 
brightest (Clump C) having a total mass of at least 2.5 $\times$ 10$^{-3}$ M$_{\odot}$ 
\citep{OGorman}. More recent ALMA measurements of Clump C  yield a dust mass 
alone of $>$ 1.2 $\times$ 10$^{-3}$ M$_{\odot}$, implying a total mass of 
$\sim$ 0.1 M$_{\odot}$ \citep{Vlemmings2017}.  These measurements imply short term, high mass loss in separate events over several hundred years.

The mass loss mechanism for red supergiants is  not understood. Leading processes include radiation pressure on grains, pulsation and convection, with convection being the most promising. The prominent arcs and dusty clumps in VY CMa   provide clues to an additional mechanism, surface activity and magnetic fields. The presence of magnetic fields is supported by the Zeeman splitting and
polarization of the masers in its circumstellar ejecta \citep{Vlemmings2002,Vlemmings2003,Vlemmings2005,Shinnaga}.  Recent ALMA continuum observations \citep{Vlemmings2017} reveal
polarized dust emission from magnetically aligned gains within 0\farcs5 of the star.

In addition to the large-scale features clearly visible in the HST images, the 
ejecta are filled with numerous small condensations and filaments, suggesting that
lesser  events have occurred. Ground-based spectroscopy 
has been limited to 0\farcs8 spatial resolution (Paper I). 
Several small dusty knots and filaments   
$\leq$ 0\farcs5 from  the star very likely represent the most recent mass loss events.
We therefore obtained {\it HST/STIS} observations with a spatial resolution of 0\farcs1 to probe  VY CMa's innermost ejecta
and the frequency of recent events.  

Our spectra of the star and the structures closest to the star yielded a surprising result. The very strong K I emission lines generated  by resonant scattering and 
the  molecular electronic spectra, assumed to form in the dusty envelope  
around the star, actually arise in the knots, not the immediate environment of the star.  
These observations also indicate that VO is actually a circumstellar molecule. In this Letter we focus on the spectra of these knots and filaments nearest the star and the identification of the circumstellar molecular transitions. We also discuss the motions of the knots, their spatial orientation, and time since ejection,  as well as the implications for the excitation of the atomic and molecular lines.  

\section{{\it HST/STIS} Observations} \label{sec:obs}

The {\it HST/STIS} observations were planned in two visits to allow for two 
different 
slit orientations to cover the small knots and filaments immediately to 
the west and east of the star with three slit positions and  to separate the 
individual knots to the south and southwest of the star with three slit positions as shown 
in Figure 1. The central star\footnote{The star is obscured by circumstellar dust, thus its spectrum is observed by scattered light. By ``central star'' we mean the bright object $r \lesssim 0.07$ arcsec.} was observed at each visit.  Slits W1 and W2 were placed over the nebulous condensations visible in Figure 1. Although there are no obviously associated optical features with ALMA Clump C, slit E1  passed over the Clump C  region and faint filaments just to the east of the star. 

We used the STIS/CCD with the G750M grating at tilts 7795{\AA} and 8561{\AA} to measure the K I emission lines at 
7665{\AA} and 7699{\AA} and the Ca II absorption triplet (8498,8542,8662{\AA}).
The K I lines, formed by resonant scattering, are the strongest emission 
features in VY CMa and are the best tracers of the gas. The strong Ca II lines
are reflected by dust. 
The 52x0.1 slit was used for the west/east knots and filaments and the 0\farcs2 slit was chosen for the knots south of the star to cover the somewhat more extended  structures. 
 At the 1.2 kpc distance of VY CMa \citep{Zhang}, 0\farcs1 is $\approx$ 120 AU.
 After acquisition, the star was centered on the slit using a peakup in the red, and the telescope was then offset to the positions shown in Figure 1. 
 The  observations were completed on 2018 January 5  and February 11.

\begin{figure}
\figurenum{1}
\epsscale{1.0}
\plotone{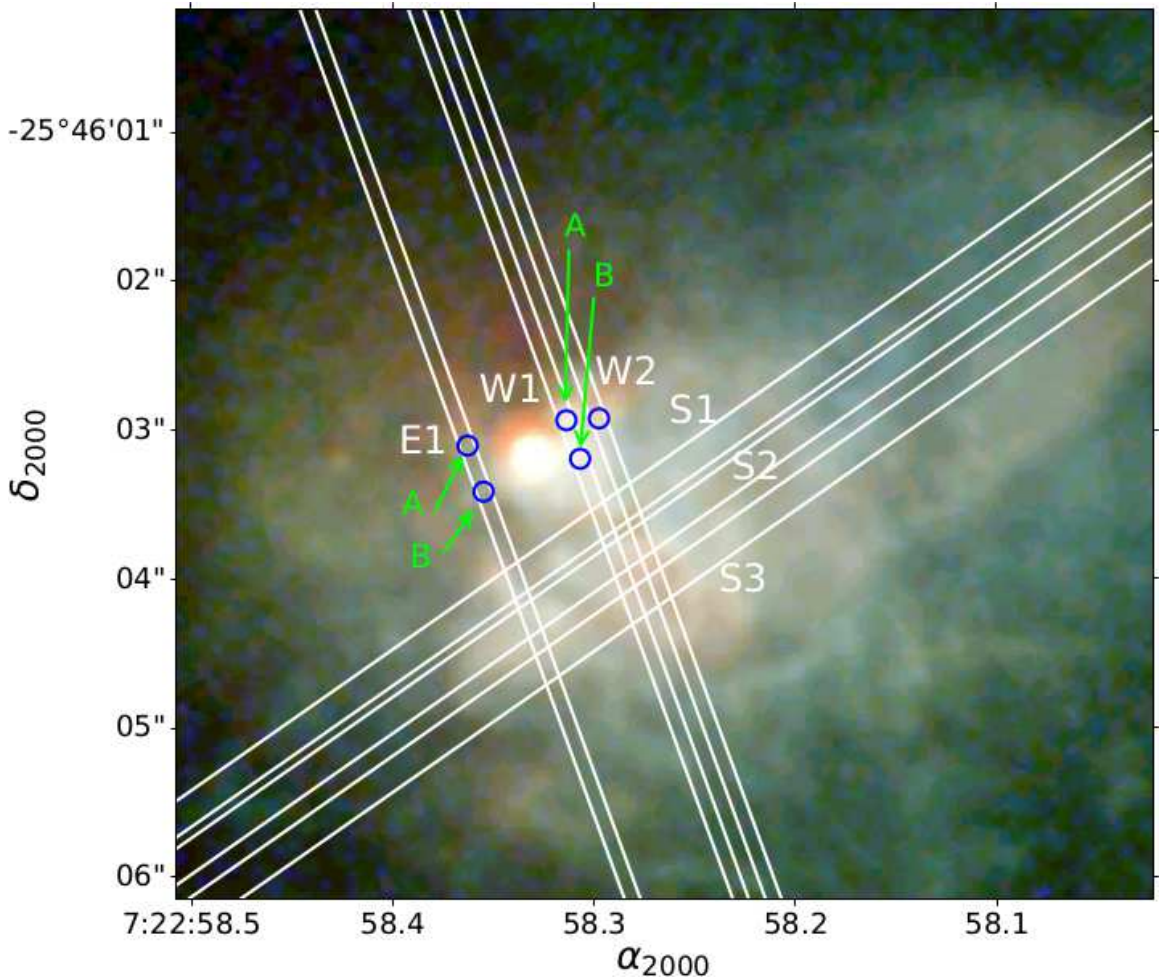}
\caption{Slit positions for the HST/STIS spectra superposed on an HST image of VY CMa. The positions for the extractions discussed in the paper are marked by 
circles.}
\label{fig:slits}   
\end{figure}

All of the spectra were processed using the sub-pixel modeling technique developed 
for the eta Car Treasury programs (Davidson 2006) to mitigate degradation of the HST's 
spatial resolution along the slit. Consequently, the  scale in the processed 
spectra is  0.0253" per pixel. Contemporary flat-field images were also 
obtained  
to correct for fringing and  were processed in the same way 
as the science images and  were normalized to be used as contemporary fringe 
flat-field images. 

\section{Spectra and Motions of the Inner Knots}  

We extracted one dimensional spectra of the visible knots in slits W1 and W2 
and at two positions in slit E1 as well as the star with a spatial width of 0\farcs1 along the slit.  Sample extracted spectra shown in 
Figure 2.  The differences between the star and W1 knot A  are immediately apparent. The molecular emission bands in the 7795{\AA} spectrum and the broad molecular emission band and atomic emission lines 
in the 8561{\AA} spectrum are unique to the three extractions west of the star. 
The spectra of knot B in W1 and the knot in W2 are very similar
to knot A with small differences in the flux level of the continuum  
due to different distances from the star. The emission bands 
are not 
present in the two extractions in E1. Nor are they present in any  
spectra extracted in slits to the south of the star.  

\begin{figure}  
\figurenum{2}
\epsscale{1.1}
\plotone{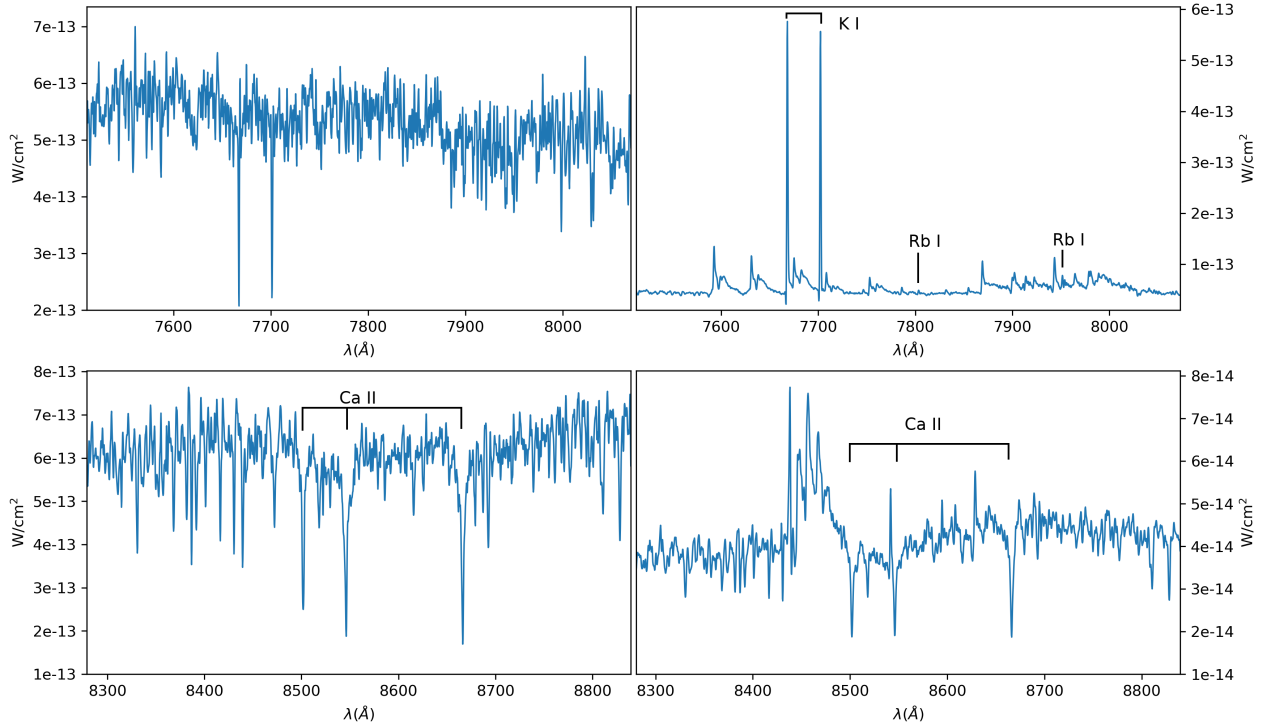}
\caption{The two upper spectra  are of  the star on the left and knot A in the W1 slit on the right  showing the strong K I emission lines and molecular bands. The bottom spectra are the star on the left and knot A in the W1 slit showing the Ca II triplet in each  and the previously unrecognized strong molecular emission band and unidentified features  in the knot. }
\label{fig:spectra}
\end{figure}

The K I emission lines are much stronger relative to the continuum level 
in W1 knots A and B and the W2 knot than on the star. The total flux in the K I emission profiles is 10 to 20 times greater than on the star. All three knots show  absorption components in the K I lines as does the star. The Rb I emission lines at 7800{\AA} and 7947{\AA} \citep{Wallerstein1971} are not detectable in the spectrum of the star.   The 8561{\AA} spectra are dominated by the broad molecular feature and the strong Ca II triplet in absorption. Numerous weaker absorption lines are present due to Fe I and Ti I.
The measured parameters of the K I emission lines, their Heliocentric velocities and the velocities of the absorption lines are summarized in Table 1 for the star and at these five positions.    

We  measured the transverse motions of the these knots and filaments following the procedure described in Paper II  using our two epochs of HST/WFPC2 images from 1999 and 2005. VY CMa was imaged in four filters, F410M, F547M, F656N, and F1042M, 
with a range of exposure times.   
The images are tightly aligned in pixel space and  blinked to identify any offset. We measure the x and y positions of the blinked images three times in each filter combination in which the knot or filament is identified. The measurements  are then combined for a weighted mean. In combination with the Doppler velocity measured from the K I lines relative to the star, we  determine each clump's   orientation, direction of motion and total space velocity  and  then estimate its age or time since ejection included in Table 1.

\begin{splitdeluxetable*}{lccccBlcccccc}   
\tabletypesize{\scriptsize}
%\rotate
\tablenum{1}
%\tablecolumns{12}
\tablecaption{Measured Fluxes, Heliocentric Velocities and Motions}
\tablewidth{0pt}
\label{tab:Meas}
\tablehead{
\colhead{Object} & 
\colhead{K I Fluxes\tablenotemark{a}} & 
\colhead{K I em Vel} & 
\colhead{K I abs Vel} & 
\colhead{Abs. Lines Vel.} &  
\colhead{Object} &  
\colhead{Trans Vel. } &  
\colhead{Direct. $\phi$} &
\colhead{Orient. $\theta$} & 
\colhead{Total Vel.} & 
\colhead{Dist.} & 
\colhead{Age} \\
\colhead{Name} &
\colhead{ergs cm$^{-2}$ s$^{-1}$} &
\colhead{km s$^{-1}$}  &
\colhead{km s$^{-1}$}  &
\colhead{km s$^{-1}$}  &   
\colhead{Name} &
\colhead{km s$^{-1}$}  &
\colhead{deg} &
\colhead{deg} &
\colhead{km s$^{-1}$}  &
\colhead{AU} &  
\colhead{yr}  
} 
\startdata 
Star  & 6.7 $\times 10^{-14}$, 3.0 $\times 10^{-14}$  & 50.8, 59.2 & 9.0, 4.7 & 67.1(21) & Star  & \nodata  & \nodata  & \nodata  & \nodata  & \nodata &  \nodata \\  
W1 knot A &  7.2 $\times 10^{-13}$, 6.5 $\times 10^{-13}$ & 42.6, 44.0 & -12.5, -10.9 & 57.7(17) & W1 knot A  & 21.6 $\pm$ 4 & -45.6 $\pm$ 6 & -28 $\pm$ 5  & 24.6 $\pm$ 4.5  & 390 & 77 $\pm$ 14 (1922) \\ 
W1 knot B  & 8.1 $\times 10^{-13}$, 8.4 $\times 10^{-13}$ & 45.8, 46.0 & -16.4, \nodata & 67.7(15)  &  W1 knot B  & 18.9 $\pm$ 6 & -14 $\pm$ 12  & -25 $\pm$ 6  & 21.5 $\pm$ 6  & 300 & 67 $\pm$ 20 (1932) \\  
W2 knot  & 5.0 $\times 10^{-13}$, 5.2 $\times 10^{-13}$ &  47.3, 46.0 & -22, -24.5 &  76.0(13) & W2 knot  & 32.6 $\pm$ 10  & -43 $\pm$ 10  &  -15 $\pm$ 5  &  32.6 $\pm$ 10  &  640 & 94 $\pm$ 28 (1905) \\
E1  A & 6.1 $\times 10^{-15}$, 1.1 $\times 10^{-14}$ & 45.4, 45.2 &  0.4, -1.5 & 81.5(15) & E1 A & 27.4 $\pm$ 5 & 50 $\pm$ 8  & -19 $\pm$ 4  &  29.0 $\pm$ 6 & 590 & 98 $\pm$ 20 (1901)  \\ 
E1 B & 5.3 $\times 10^{-15}$, 6.5 $\times 10^{-15}$ & 47.3, 44.8 & -5, -3.9 & \nodata &  E1 B & \nodata  & \nodata  & \nodata  & \nodata &  \nodata  &  \nodata \\
\enddata
\tablenotetext{a}{The fluxes here are measured in a $0\farcs1 \times 0\farcs1$ aperture. The clumps are somewhat larger.}
\end{splitdeluxetable*}

The three small condensations to the west of the star are well defined,  but the more filamentary features in slit E1 are more irregular and difficult to  measure. We were not able to measure any reproducible positions for B in E1. The results for E1 position A are based on just the highest quality/best measurements in one filter the F656 long, see Paper II. It is not surprising that these knots and filaments closest to the star are the most recent ejections. They appear to fall into two time epochs about 1920 - 1930 and $\sim$ 1900, which likely represent VY CMa's most recent active periods.

\section{The Molecular and Atomic Emission Features} 

Numerous atomic emission lines and molecular bandheads are present in the  
spectrum of VY CMa. Many of the TiO and VO transitions were observed  by \citet{Wallerstein1971,Wall86} and \citet{Wall2001}. A bandhead of ScO was also identified by Wallerstein (cited in \citet{Hyland}) and by \citet{Herbig}. We confirm  
 many of these past identifications,  as well as new molecular 
 bands, including the broad feature in the far-red spectrum. We also demonstrate that these molecules are clearly circumstellar, as opposed to being formed closer to the star in its wind. We thus present the first conclusive evidence for the identification of VO  as circumstellar.  Our identifications are summarized in Table 2 and shown in Figure 3. 

We identify VO bandheads of the B$^{4}\Pi$-X$^{4}\Sigma$ 
electronic transition. Figure 3 shows spectra of the (0,0) and (0,1) vibrational transitions. A simulation of the (0,0) spectra is also shown above the astronomical data, using pGopher \citep{Western}. The simulation reproduces the observed spectrum quite well, assuming a temperature of $\sim$ 450 K. The complexity of the spectrum arises from the presence of fine structure and P,Q, and R branch line blending in the quartet-quartet electronic transition. \citet{Wallerstein1971} identified some of these features (Table 3), but most are new assignments, based on the laboratory work of
\citet{Keenan} and \citet{Adam}. \citet{Cheung} and \citet{Adam} conducted the first accurate spectral analysis of VO, and made assignments of the fine-structure components unknown to Wallerstein. These spectra are definitive evidence for the presence of VO in circumstellar ejecta 100’s of AU from the central star. 

For TiO, the A$^{3}\Phi$-X$^{3}\Delta$ electronic band 
 is prominent with numerous sharp R-branch ($\Delta$ J = +1) bandheads  
with less intense P and Q features arising from the (0,1), (1,2) and (2,3) vibrational sequence. New identifications for TiO spectra are the R$_{2}$ (0,1) and R$_{1}$ (1,2) features, which are based on laboratory measurements of \citet{Ram}. (The subscript on the R refers to the spin-orbit component, as indicated by quantum number $\Omega$, of the lower level, where $\Omega$ = 1,2, and 3). A  simulation of the three fine structure (0,1) features is presented above the data, and reproduces the observed spectra well. The simulation assumes a temperature of $\sim$ 350 K.    

\newpage

\startlongtable   
\begin{deluxetable}{ccll}
\tabletypesize{\footnotesize}
\tablenum{2}
\tablecaption{Molecular Emission  and Atomic Emission Line Identifications}
%\tablewidth{0pc}
\label{tab:IDs}
\tablehead{
\colhead{Vacuum Wavelength {\AA}} & \colhead{Lab Vacuum Wavelength {\AA}} & \colhead{Identification} &  \colhead{Reference\tablenotemark{a}}
}
\startdata
7592.8        &  7591.5\tablenotemark{b}      &    TiO A-X R$_{3}$ (0,1) & 1, 2 \\
7631.0        & 7629.9\tablenotemark{b}       &    TiO A-X R$_{2}$ (0,1) & This work, 2 \\
7668.1         & 7667.0       &    K I              & 1    \\
7675.1       &  7673.7\tablenotemark{b}      &   TiO A-X R$_{1}$ (0,1)  & 1, 2 \\
7702.23         &  7701.10      &    K I              &  1   \\
7708.4        &  7707.1\tablenotemark{b}      &   TiO A-X $_{2}$  &  1,2  \\
7753.0        &  7751.6\tablenotemark{b}      &   TiO A-X R$_{1}$  (1,2)  &  This work, 2 \\
7786.0        &  7785.0\tablenotemark{b}       &  TiO A-X R$_{2}$ (2,3)  &  2, 3 \\
7803.4         &  7802.38      &   Rb I              &  1    \\
7831.7        &  7830.3\tablenotemark{b}      &  TiO A-X R$_{1}$  (2,3)   &   2,3  \\
7854.7        &  7853.2\tablenotemark{b}       &  VO B-X (0,0) & This work, 4, 5 \\
7869.2         & $\sim$ 7867\tablenotemark{d} & VO B-X (0,0)    &  1, 4, 5 \\
7901.3\tablenotemark{c}   & $\sim$ 7899\tablenotemark{d}  &  VO B-X (0,0) &  This work, 4,5\\
7913.8        & 7912.3\tablenotemark{e}       & VO B-X (0,0) & 1, 4, 5 \\
7922.7        & $\sim$ 7920\tablenotemark{d} & VO B-X (0,0)  & This work, 4, 5 \\
7943.4        & $\sim$ 7940\tablenotemark{d}  & VO B-X (0,0) & 1, 4, 5 \\
7951.5\tablenotemark{c}  & 7949.8       &  Rb I + VO\tablenotemark{c}   &  1  \\
7964.5        & $\sim$ 7963\tablenotemark{d}  & VO B-X (1,1) & This work, 4, 5 \\
7979.8\tablenotemark{c} & $\sim$ 7978\tablenotemark{d}  & VO B-X (1,1) &  This work, 4, 5 \\
7988.9        &  $\sim$ 7987\tablenotemark{d}  &  VO B-X (1,1) &  This work, 4, 5 \\
8438.15         & \nodata       & unknown  &   \nodata \\
8448.2        & 8446.4\tablenotemark{f}  & TiO E-X (0,0)  &  This work, 6 \\
8456.8        &  8455.6\tablenotemark{f}  &  TiO E-X (0,0) & This work, 6  \\
8467.2        & 8466.1\tablenotemark{f}       & TiO E-X (0,0)  & This work, 6 \\
8476.7        & $\sim$ 8475\tablenotemark{d}     & TiO E-X (0,0)  & This work, 6 \\
8541.8         & $\sim$ 8540\tablenotemark{d}    & VO B-X (0,1) & This work, 4, 5 \\
8594.6        & $\sim$ 8593\tablenotemark{d}    &  VO B-X (0,1)  & This work, 4, 5 \\
8608.2        & $\sim$ 8607\tablenotemark{d}    &  VO B-X (0,1) & This work, 4, 5 \\
8628.7        & $\sim$ 8627\tablenotemark{d}     & VO B-X (0,1) & This work, 4, 5 \\
\enddata
\tablenotetext{a}{References 1) \citet{Wallerstein1971}, 2) \citet{Ram}, 3) \citet{Wall2001}, 4) \citet{Keenan} 5) \citet{Adam} 6) \citet{Simard}}
\tablenotetext{b}{Approximate R-branch bandhead} 
\tablenotetext{c}{Blended feature}
\tablenotetext{d}{Blended branches}
\tablenotetext{e}{Q-branch bandhead}
\tablenotetext{f}{Estimate of Q-branch maximum} 
\end{deluxetable}

\begin{figure}  
\figurenum{3}
\epsscale{1.4}
\plottwo{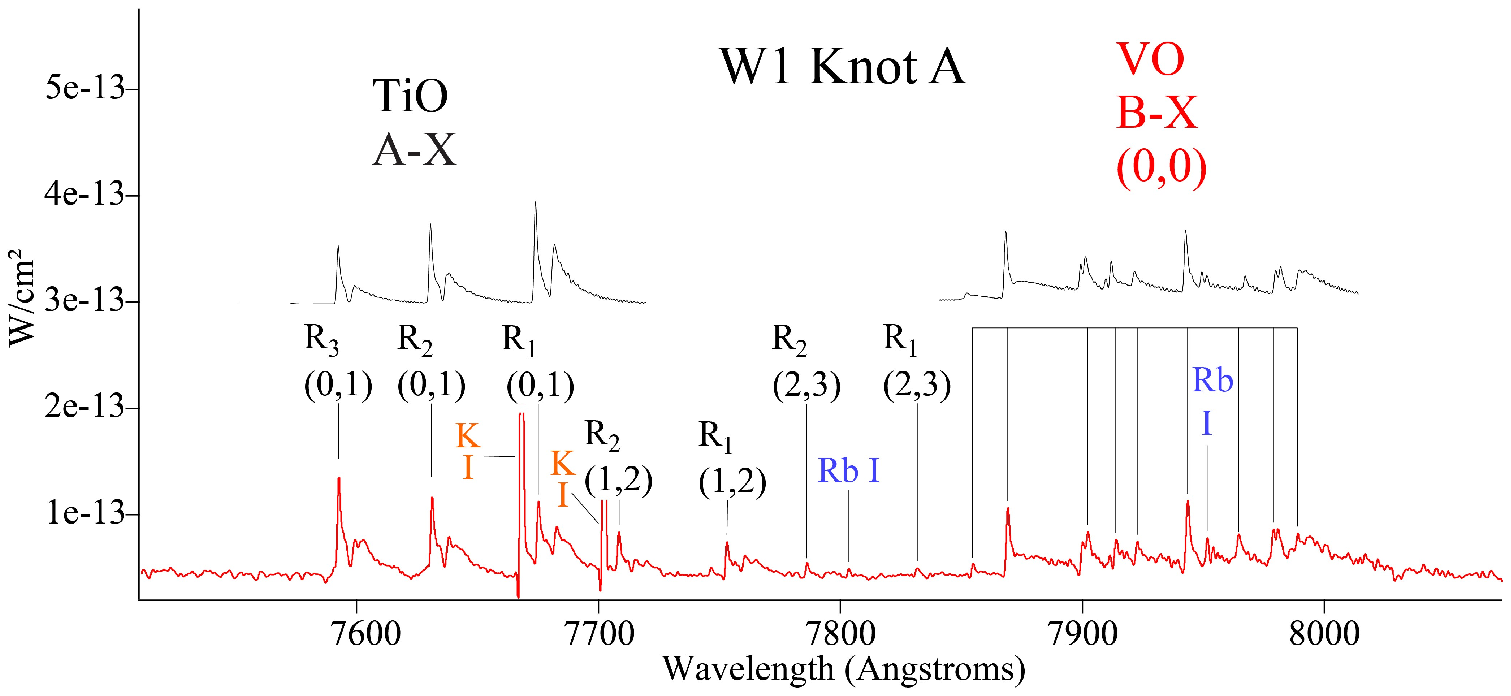}{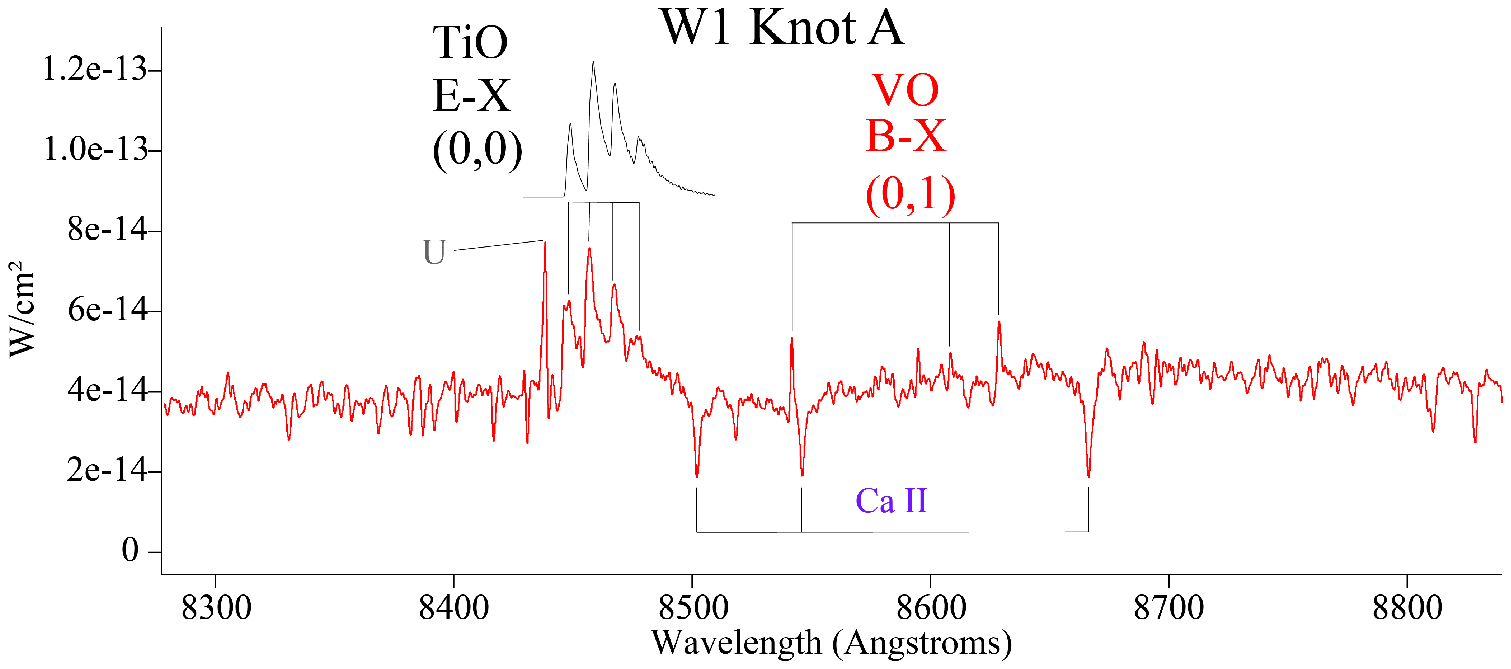}
\caption{The molecular identifications from Table 2 shown on the red and far red spectra. Simulations are shown above the stronger transitions. The \ion{K}{1} lines are truncated.}
\end{figure}

In addition to the A-X electronic transition, the E$^{3}\Pi$-X$^{3}\Delta$  transition of the (0,0) vibrational band of TiO is observed for the first time in 
VY CMa.  As measured in the laboratory by \citet{Simard}, this band consists of 
four prominent peaks, arising from strong Q-branches blended  with P and R-branch transitions, as shown by the simulation above the observed data. The E-X transition has been previously reported in the peculiar red nova merger remnants V4332 Sgr and V1309 Sco \citep{Tylenda,Kaminski15}.  
 The TiO E-X spectrum was best reproduced assuming T $\sim$ 250 K. Note that the excited A and E states lie ∼ 13,000 cm$^{-1}$ and ∼ 11,800 cm$^{-1}$ higher in energy above the ground states. Despite its designation, the E state actually lies lower in energy by about 1250 cm$^{-1}$.

TiO$_{2}$  emission has been mapped within 1\arcsec  around the central star  by \citet{debeck} from observations with ALMA with $\sim$ 0\farcs2 resolution.  
The morphology of 
the TiO$_{2}$ emission with LSR velocities greater than 44 km s$^{-1}$  
overlaps the  strong TiO (and K I) emission in the knots to the west of the 
star. The TiO A-X bandheads in W1 knot A have a mean Heliocentric velocity of 53.4 $\pm$ 6.0 km s$^{-1}$, 68 km s$^{-1}$ LSR.  This spatial and kinematic overlap suggests that the  
TiO$_{2}$ emission may originate from the same dusty condensations. But 
a comparable correspondence  does not apply to the region just east of the star. The velocities of the TiO$_{2}$ emission there are blueshifted with an LSR velocity range of 28 -43 km s$^{-1}$ compared to the LSR velocity ($\sim$ 60  km s$^{-1}$)  of the visible filaments to the east, but their spectra lack 
TiO  emission.  

One feature near 8438{\AA} remains unidentified. The line profile is quite sharp, suggesting an atomic origin. However, Herbig (1974) explained how a complex structure in an electronic transition could be shrunk down to a single apparent  line in VY CMa due to the crowding of the rotational structure near the bandheads. Possible wavelength agreements with Sc I and Ca I/Ca I] all involve unlikely high level transitions.

\section{A Basic Astrophysical Question -- Excitation of the Strong K I and Molecular Emission in the Ejecta}  

Evidently the extraordinary \ion{K}{1} doublet  and the strong molecular emission,  
 originate primarily in a few small 
diffuse clumps  0\farcs3  west of the central star, unresolved in  
 ground-based spectra. Our results  significantly alter the existing models.    
In this Letter, we offer only a simplified outline of the problem\footnote{We do not discuss various semi-obvious caveats -- e.g., the dependence on viewing 
direction, whether brightness might not imply  a maximum in the gas density, etc.}. 

\ion{K}{1} is observed throughout the the extended ejecta via resonant scattering (Paper 1), but
 the strong molecular emission is observed only in the clumps to the west of the
star.  
As our discussion below suggests,  radiative excitation and resonant scattering
are  feasible for the strong \ion{K}{1} doublet in the clumps.  On the other hand, gas temperatures are probably  important for the molecular emission. Our derived temperatures of 300 -- 400 K   are  close to the expected dust-grain temperatures and radiation-density temperatures in the clumps.   Excitation by shocks or MHD waves may 
be important for the molecules, but shocks have at least one disadvantage: 
    the observed temperatures require low shock speeds of the order 
    of 4 km s$^{-1}$, inefficient for energy transport. Thus our measurements support radiative  
excitation of the molecules which originate in the clumps, but collissional process may play a role.  

The \ion{K}{1} doublet is VY CMa's strongest emission feature, 
and it leads to a remarkable unpredicted result described  
below.   With a total luminosity of the order of 4 $L_\odot$ in just 
two narrow lines (Paper I), it greatly exceeds the \ion{K}{1} features  
known in other evolved cool stars \citep{BL76,PL94,GM96}.  
 As explained  in 
Paper I, resonance scattering of continuum photons is the only 
likely production mechanism.  Paper I emphasized that this 
process is a zero-sum game; we must explain why the spectrum has 
no strong absorption feature comparable to the observed emission   
-- i.e., like the absorption part of a pure-scattering P Cygni profile.  
In Appendix A2 of Paper I,  we explained how the absorption can be hidden 
if the dust-formation zone at $r \gtrsim 80$ AU is very inhomogeneous.  
That scenario is too lengthy to repeat here, but a crucial point is 
that most escaping photons -- roughly 2\% of the star's output in the 
relevant wavelength range -- follow multiply-scattered paths between 
opaque condensations.  In the simplest viable model, emergent 
\ion{K}{1} would have a mottled appearance with a diameter of  
the order of 600 AU ($\sim$ 500 mas). 

However, the HST/STIS data now show that most of the \ion{K}{1} emission 
comes from a few condensations  -- for instance,  
Knot W1-A produces roughly 10\% of the total \ion{K}{1} flux  
associated with the ``star image'' in ground-based spectra (Paper I). 
Relatively little \ion{K}{1} emission appears in the central object  
($r \lesssim 80$ mas or 100 AU). 
In the only model that appears realistic to us, the small clumps  must  
``see''   remarkably clear holes in the circumstellar dust shell.
We expect to publish a detailed analysis later, and here we  
only sketch the reasoning for W1 knot A.  
(1) With a diameter of 200 AU or less, located 400 AU from the star, 
the knot intercepts 1\%  to 2\% of the star's light if there is no dust 
between them.  (2) With $L \sim 3 \times 10^5 \; L_\odot$ and $T \sim 3500$ K,
the star radiates about 25 $L_\odot$ per {\AA}  around 7700 {\AA};
so the knot intercepts 0.5 $L_\odot$ per {\AA} or less.
(3) Each of the \ion{K}{1} lines from knot W1-A  has a luminosity 
of roughly 0.3 $L_\odot$, corrected for interstellar extinction.  
(4) Therefore, if other conditions are favorable, then the knot's 
observed \ion{K}{1} brightness can be attained if the incident continuum 
in wavelength interval $\Delta\lambda \gtrsim 0.6$ {\AA} is converted 
via resonance scattering in the knot,  required for  
the two  \ion{K}{1} lines.  The corresponding velocity width is of the order 
of 30 km s$^{-1}$, which appears possible with an allowable velocity 
dispersion in a knot and the \ion{K}{1} oscillator strengths. 
The main result of this exercise is that the path from the star 
to each bright knot must be nearly free of dust; a 50\% extinction 
loss would increase the needed $\Delta\lambda$ to an implausible value.    

This is a remarkable result, since dust is expected to have optical 
depths $\tau \sim 50$ along typical radial paths (Paper I).  
According to the above analysis, $\tau \lesssim 0.5$ in a roughly 
$30^\circ$-wide cone toward each bright 
knot.  Thus, during roughly a century since the knots were ejected, 
the star ejected practically no material in those directions -- even 
though it presumably rotates. The strong \ion{K}{1} have been reported since 1958 \citep{Wallerstein58}.  We suspect that this may be explained 
 in terms of the statistics of size scales and time 
scales in the outflow structure;  but it was not expected, and it gives   
an unpredicted constraint on the outflow physics.  

As we emphasized earlier, the mass loss mechanisms for RSGs are   
uncertain;  and VY CMa is arguably the most significant example because  
it is  extreme in several respects.   Its localized sources of 
molecular emission and scattered atomic lines, reported above,   
were not predicted; and they imply major gaps or holes in the   
outflow structures formed by large-scale stellar activity. 
Related processes may occur in more normal RSGs.  High-resolution
imaging of $\alpha$ Ori has  revealed small condensations and filaments within
1$\arcsec$ \citep{Kervella}.   Future theoretical models of active regions 
must account for these observed density contrasts and size scales.

\acknowledgements
R. Humphreys thanks Kris Davidson and George Wallerstein for useful comments and discussion. This work was supported by NASA through grant GO-15076 (P.I. R. Humphreys) from the Space Telescope Science Institute and National Science Foundation grant  AST-1515568 (P.I. L. Ziurys). 

\vspace{5mm}
\facilities{HST(STIS)}

%% Similar to \facility{}, there is the optional \software command to allow 
%% authors a place to specify which programs were used during the creation of 
%% the manusscript. Authors should list each code and include either a
%% citation or url to the code inside ()s when available.

\end{document}